# Improved Twitter Sentiment Prediction through 'Cluster-then-Predict Model'

[1] **Rishabh Soni,** [2] **K. James Mathai**

[1] PG Scholar, Department of Computer Engineering and Applications, NITTTR Bhopal,
Bhopal, Madhya Pradesh, India

[2] Associate Professor, Department of Computer Engineering and Applications, NITTTR Bhopal,
Bhopal, Madhya Pradesh, India

**Abstract** - Over the past decade humans have experienced exponential growth in the use of online resources, in particular social media and microblogging websites such as Facebook, Twitter, YouTube and also mobile applications such as WhatsApp, Line, etc. Many companies have identified these resources as a rich mine of marketing knowledge. This knowledge provides valuable feedback which allows them to further develop the next generation of their product. In this paper, sentiment analysis of a product is performed by extracting tweets about that product and classifying the tweets showing it as positive and negative sentiment. The authors propose a hybrid approach which combines unsupervised learning in the form of K-means clustering to cluster the tweets and then performing supervised learning methods such as Decision Trees and Support Vector Machines for classification.

## 1. Introduction

Companies traditionally rely on interviews, questionnaires and surveys to gain insight through customer feedback about their products and the company itself. These traditional methods are often extremely time consuming and expensive. They does not always return the desired results that the companies were looking for due to environmental factors and poorly designed survey instrument. In this paper, the results of sentiment analysis is carried out to get public perception of the company 'Apple'.

The Twitter Streaming API was used to obtain real-time tweets as text about the company for sentiment prediction. The real-time collection of tweets at three stages i.e. after the initial announcement of the product, at the product launch and after the product has been launched; have helped in analyzing the sentiment towards the company.

Decision trees such as Classification and Regression Trees (CART) and Random Forests were used for classifying tweets sentiments as positive or negative. They also provided visual insight as which words as features are most prominent in swaying the perception of the company. This would help the company to identify the services and products which satisfies the customers.

There are different methods for classifying tweets such as Support Vector Machines (SVM), Logistic Regression, Classification and Regression Trees (CART), Random Forests etc. The author(s) propose a hybrid method by which:

- The tweets are clustered based on the words they contain by applying K-means clustering on the data, and
- Train the data by applying Random Forest on each of the subset of tweets for classification.

The predicted results of this research when compared with the results of traditional approaches such as CART, Random Forests and SVM, have shown better accuracy.

This method can also be called as **'cluster-then-predict Model'** because in this model, firstly the similar type of tweets are clustered depending upon the sentiment of words they contain and then train the model for prediction. The accuracy of the results can be shown using a confusion matrix. Thus, the proposed hybrid approach shows that prediction with such a method results in better accuracy.

## 2. Related Work

Due to the widespread presence of internet and growth of social media websites such as Facebook and Twitter, online social network analysis has become a hot research topic. Twitter has an easy to use API that provides textual data for a variety of purposes. In paper [1], authors Alexander Pak et al. focusses on twitter for the task of sentiment analysis. They perform linguistic analysis of the collected corpus and explains the discovered phenomena. They build a classifier based on the multinomial Naïve Bayes classifier that used N-gram and POS-tags as features.

Stock prediction using twitter corpora is a major research area and provides direction of work. Many researches have been carried out in this aspect. In the study [2], the authors Vu, Tien-Thanh et al., harnesses features from twitter messages to capture public mood related to four Tech companies for predicting the daily up and down price movements of these companies' NASDAQ stocks. In the study [3], author Linhao Zhang examines the effectiveness of various machine learning techniques on providing a positive or negative sentiment on a twitter corpus. The author have applied supervised machine learning techniques like SVM, Naïve Bayes, maximum Entropy etc. and compared the results. They also looked for correlation between twitter sentiments with stock prices and determined which words in tweets correlate to changes in stock prices by doing a post analysis of price change and tweets.

Due to the varied nature of textual information available on twitter, various analyses like public opinion on a television show [4], sentiment analysis on movie reviews [5], stock prediction [2][3][6], automatic crime prediction [7] and election results prediction [8] have been carried out. In this research, authors have used sentiment analysis technique for finding out public perception about the company so that it can devise its future strategy for itself.

## 3. Proposed Work

Authors proposes a hybrid approach which comprises of both unsupervised and supervised learning for predicting the sentiment (refer section 3.3 & 3.4). After obtaining the data(refer section 3.1) and preprocessing it, K-means clustering is performed to form clusters of tweets data points, such that similar tweets (based on the words they contain) gets clustered into one cluster. The unsupervised learning stage is mostly used in conjunction with feature extraction stage. After performing K-means clustering i.e. unsupervised training on the data; supervised training is done on the same data. In it, the data is divided into training and testing sets, each containing 70% and 30% of the data, respectively. After that, proposed learning model is applied to each of the training data set individually. There are various supervised learning methods for building the model, but author(s) found that the Random Forests algorithm is best suited for the problem because it provides best trade-off between accuracy, interpretability and execution time.

Author(s) proposes supervised learning pipeline as shown by the diagram below, clustering is done in conjunction with feature extraction stage, to implement 'cluster-then-predict Model' for better prediction results.

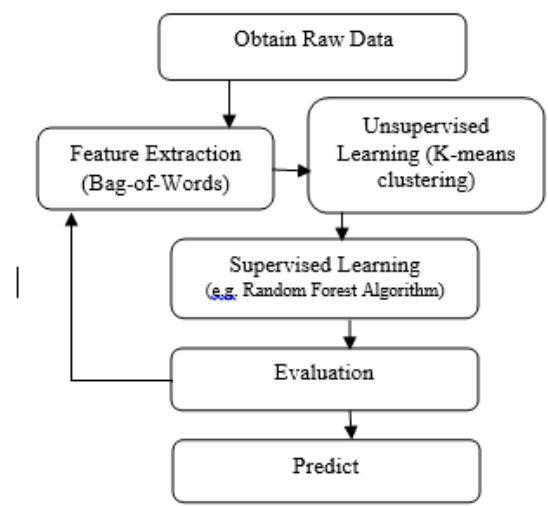

Fig. 1. 'Cluster-then-predict Model'- Proposed hybrid learning pipeline

### 3.1 Obtaining Raw Data

To verify sentiment prediction through proposed 'Cluster-then-predict Model', Twitter's tweets have been used as text data. Twitter data that is publically available were collected by Twitter API. Twitter's streaming API service is used to store real-time tweets. Python's API named Tweepy [9] have been used to implement streaming API of Twitter. It provides libraries to collect streaming twitter data. The incoming tweets were stored in CSV (Comma Separated Values) file format in real-time by importing Python's CSV library functions.

For this research, 1200 tweets were collected for the company 'Apple' for analysis. After the tweets have been saved in the CSV file, the sentiment value for each tweet is assigned for the sake of supervised learning. This sentiment value assigned by workers on job by Amazon's Mechanical Turk (MTurk) [11] is used in the proposed work.

### 3.2 Feature Extraction

The feature extraction process is derived from Bag-of-words [10] approach. (Here the text is represented as a bag of its words.) The frequency of occurrence of each word is used as a feature for training the classifier- Radom Forest Algorithm. Redundant and sparse data is also removed from initial raw data, so that the reduced set of features takes less time in algorithms and also it reduces overfitting of the training set.

### 3.3 Unsupervised Learning: K-means Clustering

After forming the bag-of-words from twitter corpus, K-means clustering has been performed on it, to partition the dataset into K clusters. This action, partitions the tweets according to the words they contain. This classifies the tweets having the similar words will get one cluster. This process will in turn lead to a better prediction on the test dataset.

The K-means clustering follows a simple and easy way to classify a given dataset through a certain number of cluster 'k', fixed prior to implementing it. The main steps of K-means clustering can be shown by the Table 1 below:

Table 1: K-Means Clustering

| *K*-Means Clustering Algorithm |
|---|
| 1. Specify the desired number of clusters 'k' |
| 2. Randomly assign each data point to a cluster |
| 3. Compute cluster centroids |
| 4. Reassign each point to closest cluster centroid |
| 5. Re-compute cluster centroids |
| 6. Repeat 4 & 5 until no improvements are made |

This algorithm aims at minimizing an *objective function*, in this case a squared error function. The objective function can be represented as follow:

$$J = \sum_{j=1}^{k} \sum_{i=1}^{n} \left\| x_i^{(j)} - c_j \right\|^2$$

where $\|x_i^{(j)} - c_j\|^2$ is a chosen distance measure between a data point $x_i^{(j)}$ and the cluster center $c_j$, is an indicator of the distance of the *n* data points from their respective cluster centers.

### 3.4 Supervised Learning: Decision Trees

Decision trees are one of the most widely used machine learning algorithms much of their popularity is due to the fact that they can be adapted to almost any type of data. They are a supervised machine learning algorithm that divides its training data into smaller and smaller parts in order to identify patterns which can be used for classification. The knowledge is then presented in the form of logical structure similar to a flow chart that can be easily understood without any statistical knowledge. The algorithm is particularly well suited to cases where many hierarchical categorical distinctions can be made.

Here, the Random Forest algorithm is used to classify the tweets as positive or negative. By plotting the model, prominent keywords generated can be depicted which sway public perception.

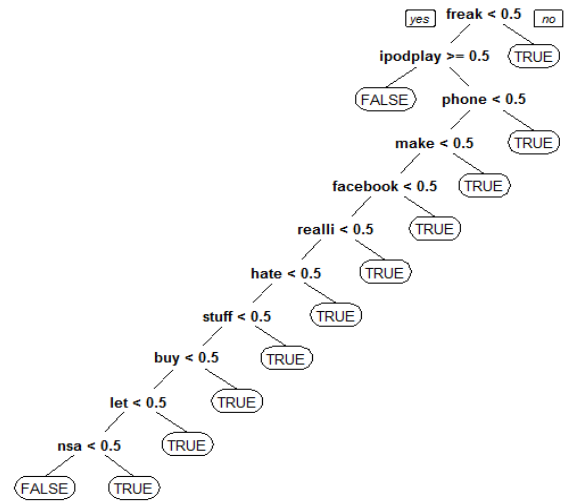

Fig. 2. Plotting Random Forest Tree

Using decision trees such as Random Forests makes solution more interpretable. For example, in this case, if the word *freak* appears in the tweet, then it is most likely to be classified into TRUE, or showing negative emotion towards the company.

### 3.5 Evaluation

Apart from using Random Forest Algorithm, many other classification algorithms can be used such as CART, Support Vector Machines (SVM), logistic regression, etc. They were also evaluated based on the parameters such as accuracy, area under ROC curve and interpretability for prediction the sentiment.

### 3.6 Prediction

After selecting and evaluating the proposed 'Cluster-then-predict Model', predictions on the test set was done. The predicted and actual results can then be compared using a *confusion matrix* shown below.

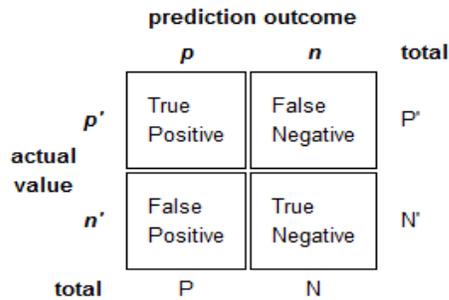

Fig. 3.  Confusion Matrix

In this paper the accuracy of the model has been found out by the confusion matrix by the formula given below:

$$\text{Accuracy} = \frac{TP + TN}{TP + FN + FP + TN}$$

Another way to examine the performance of the model is ROC (Receiver Operator Characteristic) graph. A ROC graph is a plot with the false positive tare on the *X*-axis and the true positive rate on the *Y*-axis. The ROC curves for k=2 clusters are shown in following figure:

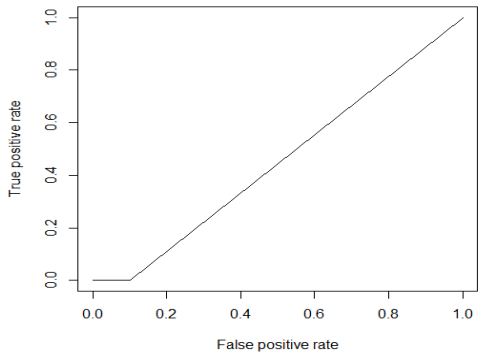

Fig. 4.  ROC curve for cluster 1

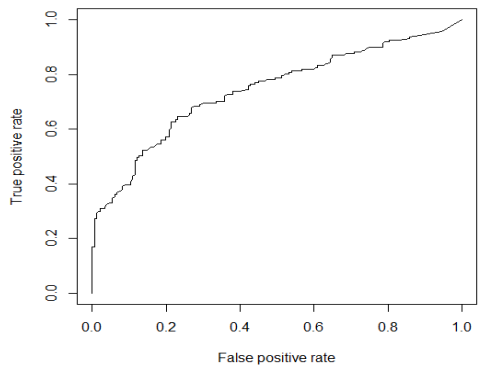

Fig. 5.  ROC curve for cluster 2

The area beneath an ROC curve can be used as a measure of accuracy.

For the proposed hybrid model, the Area under Curve (AUC) and prediction accuracy derived from the confusion matrix is higher as compared to the standalone models, as well as baseline accuracy, which is 54.23%.

Below is a short review table summarizing the evaluation of different approaches:

Table 2: Evaluation of Various Classification Algorithms

| Technique | Parameters | | |
|---|---|---|---|
| | Accuracy | Area Under Curve(AUC) | Interpretability |
| Proposed hybrid approach | 72.33% | 0.7493671 | High |
| SVM | 70.33% | 0.6934278 | Low |
| CART | 64.97% | 0.6441454 | High |
| Random forest | 70.62% | 0.7473476 | High |
| Logistic Regression | 64.40% | 0.6441454 | Low |

The table here shows that proposed "cluster-then-predict" approach has better Accuracy and AUC for sentiment prediction. It also has better interpretability so that companies can gain insights to make better decisions in the future.

## 4. Conclusions

This paper presents a hybrid mechanism- 'Cluster-then-predict Model' to improve accuracy of predicting twitter sentiment. The possibility of combining both unsupervised learning and supervised learning, in the form of K-means clustering and Random Forest, respectively performed better, than various supervised learning algorithms, such as CART, SVM, logistic Regression, etc.

The architecture of this model is scalable, so it can accommodate large amount of twitter text data. This hybrid model will perform even better when the raw data is very large and diversified.

In future work, author(s) will try to include sentiment from emoticons, make the classification multi-class i.e. showing strongly negative, negative, neutral, positive or strongly positive emotions.

**Mr. Rishabh Soni** is B.E. (Computer Science and Engineering) from Rajiv Gandhi Proudyogiki Vishwavidyalaya, Bhopal. Currently pursuing his M.TECH (Computer Technology and Applications) from NITTTR Bhopal.

**Dr. K. James Mathai** is working as Associate Professor in Department of Computer Engineering and Applications, NITTTR, Bhopal.